\documentclass{ws-procs9x6-cpt19}

\usepackage{slashed}
\usepackage{physics}
\begin{document}

\newcommand{\refeq}[1]{(\ref{#1})}
\def\etal {{\it et al.}}

\title{Lorentz Violation and Partons}

\author{Nathan Sherrill}

\address{Department of Physics, Indiana University, Bloomington, IN 47405, USA}

\begin{abstract}
A parton-model description of high-energy hadronic interactions 
in the presence of Lorentz violation is presented. 
This approach is used to study lepton--hadron and hadron--hadron interactions 
at large momentum transfer. 
Cross sections for deep inelastic scattering and the Drell--Yan process 
are calculated at first order for minimal and nonminimal Lorentz violation. Estimated bounds are placed 
using existing LHC and future US-based Electron--Ion Collider data. 
\end{abstract}

\bodymatter
\section{Introduction}
The effective-field-theory framework 
to search for potential signals of Lorentz and CPT violation 
is known as the Standard-Model Extension (SME).\cite{ck,akgrav} 
Despite numerous bounds that have been placed,\cite{datatables} 
the QCD sector of the SME is relatively unexplored 
both experimentally and phenomenologically. 
This work, 
which is based on Ref.~\refcite{parton},
describes a method for accessing some of this terrain 
from high-energy hadronic processes.


\section{Description}
In the presence of spin-independent Lorentz-violating effects on quarks, 
the dispersion relation is modified from the conventional one $k^2 = m^2$ 
and reads\cite{fermionLV}
\begin{align}
\label{eq:disprel}
k^2 -F\left(k^\mu,m, \hat{\mathcal{Q}}\right) &\equiv \widetilde{k}^2  = m^2 .
\end{align}
In general, 
the function $F$ depends on the coefficients for Lorentz violation 
$\hat{\mathcal{Q}}$, 
which are associated with operators of arbitrary mass dimension. 
To lowest order in electroweak interactions, 
partons (quarks) of momentum $k^\mu$ 
may be approximated as on-shell and massless, 
$k^2 = 0$, 
leading in the conventional case 
to the parameterization of the parton momentum 
as a fraction of the hadron momentum, 
$k^\mu = \xi p^\mu$. 
This parameterization, 
however, 
is inconsistent with Eq.\ \refeq{eq:disprel}; 
instead, 
the choice $\widetilde{k}^\mu = \xi p^\mu$ satisfies the conditions of interest. 
The implications of this parameterization are studied 
for minimal $c$-type and nonminimal $a^{(5)}$-type quark coefficients, 
which were first studied in Refs.\ \refcite{LVdis,LVgauge}, respectively, 
for the process of deep inelastic scattering (DIS).

\subsection{Deep inelastic scattering}
The DIS process $l + H \rightarrow l' + X$ describes a lepton $l$ 
delivering a large momentum transfer $-q^2$ 
upon scattering from a hadron $H$ producing a hadronic final state $X$. 
In the DIS limit $-q^2 \rightarrow \infty$ with $x = -q^2/(2 p\cdot q)$ fixed, 
where $p$ is the hadron momentum, 
the on-shell and massless limit of Eq.\ \refeq{eq:disprel} 
yields the differential cross section 
\begin{equation}
\label{eq:disxsecctheory}
\frac{d\sigma}{dx dy d\phi} = \frac{\alpha^2y}{2 q^4}\sum_f e_f^2\frac{1}{-\widetilde{q}_f^2} L_{\mu\nu}H_f^{\mu\nu}f_f(\widetilde{x}_f,c_f^{pp}),
\end{equation}
where
\begin{align}
\label{eq:omegafc}
&L_{\mu\nu}H_f^{\mu\nu} = 8 \left[2(\widehat{k}_f\cdot l)(\widehat{k}_f\cdot l') + \widehat{k}_f\cdot(l-l')(l\cdot l')   - 2(l\cdot l')c_f^{\widehat{k}_f\widehat{k}_f}\nonumber \right. \\
& \left.+ 2(\widehat{k}_f\cdot l)\left(c_f^{\widehat{k}_fl'} +c_f^{l'\widehat{k}_f} - c_f^{l'l'} \right) + 2(\widehat{k}_f\cdot l')\left(c_f^{\widehat{k}_fl} +c_f^{l\widehat{k}_f} + c_f^{ll}\right)\right],
\end{align}
with $\widehat{k}_f^\mu \equiv \widetilde{x}_f(p^\mu-c_f^{\mu p})$, $q_f^\mu = (\eta^{\mu\nu} + c_f^{\mu\nu})q_\nu$, 
and $y=(p\cdot q)/(p\cdot k)$. 
The shifted Bjorken variable $x_f = x\left(1+2c_f^{qq}/q^2 \right)  + (x^2/q^2)\left(c_f^{pq} + c_f^{qp}\right)$ to first order in $c_f^{\mu\nu}$. 
The cross section for the quark $a^{(5)}$-type coefficients 
is given by Eq.\ (61) in Ref.\ \refcite{LVgauge}, 
which is also consistent with Eq.\ \refeq{eq:disprel} 
with the proton coefficients $a_p^{(5)\mu\alpha\beta} = 0$. 
Results for both cases are also consistent with the electromagnetic Ward identity 
and the operator product expansion.

\subsection{The Drell--Yan process}
Applying the parameterization Eq.\ \refeq{eq:disprel} to the Drell--Yan process 
$H_1+ H_2 \rightarrow l_1 + l_2 + X$ 
gives the total cross section in the center-of-mass frame 
for the $c$-type coefficients:
\begin{align}
\label{eq:DYsigmac}
&\sigma = \frac{2\alpha^2}{3s}\frac{1}{Q^4}\int d\Omega_l \frac{d\xi_1}{\xi_1}\frac{d\xi_2}{\xi_2}\sum_f e_f^2\left[(\widetilde{k}_{1}\cdot l_{1})(\widetilde{k}_{2}\cdot l_{2}) +  (\widetilde{k}_{1}\cdot l_{2})(\widetilde{k}_{2}\cdot l_{1})  \right.\nonumber\\
& \left. + (\widetilde{k}_{1}\cdot l_{1})\left(c_f^{\widetilde{k}_{2}l_{2}} + c_f^{l_{2}\widetilde{k}_{2}}\right)+ (\widetilde{k}_{1}\cdot l_{2})\left(c_f^{\widetilde{k}_{2}l_{1}} + c_f^{l_{1}\widetilde{k}_{2}}\right) + (\widetilde{k}_{2}\cdot l_{1})\left(c_f^{\widetilde{k}_{1}l_{2}} + c_f^{l_{2}\widetilde{k}_{1}}\right) \right. \nonumber \\
&\left. + (\widetilde{k}_{2}\cdot l_{2})\left(c_f^{\widetilde{k}_{1}l_{1}} + c_f^{l_{1}\widetilde{k}_{1}}\right) - (\widetilde{k}_{1}\cdot \widetilde{k}_{2})\left(c_f^{l_{1}l_{2}} + c_f^{l_{2}l_{1}}\right) - (l_{1}\cdot l_{2})\left(c_f^{\widetilde{k}_{1}\widetilde{k}_{2}} + c_f^{\widetilde{k}_{2}\widetilde{k}_{1}}\right) \right] \nonumber\\
&\times\left(f_{f}(\xi_{1},c_f^{p_1p_1})f_{\bar{f}}(\xi_{2},c_f^{p_2p_2}) + f_{f}(\xi_{2},c_f^{p_2p_2})f_{\bar{f}}(\xi_{1},c_f^{p_1p_1})\right).
\end{align}
where $\widetilde{k}^\mu_i = \xi_ip_i^\mu$ for $i = 1, 2$. 
As with DIS, 
the $a^{(5)}$-type quark coefficients yield a similar expression. 
The Ward identity is also satisfied in both cases.

\section{Results}
Using existing data from the LHC\cite{CMS} 
and pseudodata for the future Electron--Ion Collider (EIC),\cite{eicLS} 
the best estimated limits for the equivalent coefficient combinations 
in the Sun-centered celestial-equatorial frame 
are shown in Table\ \ref{tab:Table1}.
\begin{table}[h!]
\centering
\tbl{Comparison of $u$-quark coefficients between the EIC and LHC. 
Bounds are reported in units of $10^{-5}$ and $10^{-6}$\text{GeV}$^{-1}$ for minimal and nonminimal coefficients, respectively.}
{\begin{tabular}{ccc}\toprule
 &EIC & LHC \\ \hline
 $|c_{u}^{XX}-c_{u}^{YY}|$& 0.74 & 15 \\
 $|c_{u}^{XY}|$ & 0.26 & 2.7 \\ 
 $|c_{u}^{XZ}|$ & 0.23 & 7.3 \\ 
$|c_{u}^{YZ}|$ & 0.23  & 7.1 \\ \hline
 $|a^{(5)TXX}_{\text{S}u} - a^{(5)TYY}_{\text{S}u} |$& 0.15 \text{GeV}$^{-1}$& 0.015 \text{GeV}$^{-1}$\\ 
  $|a^{(5)TXY}_{\text{S}u} |$& 0.12 \text{GeV}$^{-1}$& 0.0027 \text{GeV}$^{-1}$\\ 
$|a^{(5)TXZ}_{\text{S}u}|$ & 0.13 \text{GeV}$^{-1}$& 0.0072 \text{GeV}$^{-1}$\\ 
 $|a^{(5)TYZ}_{\text{S}u} |$&  0.13 \text{GeV}$^{-1}$& 0.0070 \text{GeV}$^{-1}$\\ 
 \botrule
\end{tabular}}
\label{tab:Table1}
\end{table}
These results suggest the $c$-type coefficients are more sensitive to lepton--hadron colliders, 
whereas $a^{(5)}$-type coefficients are more sensitive to hadron--hadron colliders.

\section*{Acknowledgments}
This work was supported by the United States Department of Energy under grant no.\ {DE}-SC0010120, the Indiana University Space Grant Consortium, and by the Indiana University Center for Spacetime Symmetries.

\end{document}